%% file: 6819phot.tex
\documentclass[preprint]{aastex}

\shorttitle{A Photometric Study of NGC 6819}
\shortauthors{Anthony-Twarog, Deliyannis, Twarog}

\begin{document}


\title{A $uvbyCa$H$\beta$ Analysis of the Old Open Cluster, NGC 6819 \thanks{WIYN Open Cluster study LXI}}


\author{Barbara J. Anthony-Twarog}
\affil{Department of Physics and Astronomy, University of Kansas, Lawrence, KS 66045-7582, USA}
\email{bjat@ku.edu}

\author{Constantine P. Deliyannis}
\affil{Department of Astronomy, Indiana University, Bloomington, IN 47405-7105 }
\email{cdeliyan@indiana.edu}

\and

\author{Bruce A. Twarog}
\affil{Department of Physics and Astronomy, University of Kansas, Lawrence, KS 66045-7582, USA}
\email{btwarog@ku.edu}




\begin{abstract}
NGC 6819 is a richly populated, older open cluster situated within the Kepler field. A CCD survey of the cluster on the 
$uvbyCa$H$\beta$ system, coupled with proper-motion membership, has been used to isolate 382 highly probable, single-star
unevolved main-sequence members over a 20$\arcmin$ field centered on the cluster. From 278 F dwarfs with high precision
photometry in all indices, a mean reddening of $E(b-y)$ = 0.117 $\pm$ 0.005 or $E(B-V)$ = 0.160 $\pm$ 0.007 is derived, 
where the standard errors of the mean include both internal errors and the photometric zero-point uncertainty. With 
the reddening fixed, the metallicity derived from the same 278 stars is [Fe/H] = -0.116 $\pm$ 0.101 from $m_1$ and 
-0.055 $\pm$ 0.033 from $hk$, for a weighted average of [Fe/H] = -0.06 $\pm$ 0.04, where the quoted standard errors of the mean values include 
the internal errors from the photometric scatter plus the uncertainty in the photometric zero points. If metallicity 
is derived using individual reddening values for each star to account for potential reddening variation across the face 
of the cluster, the analogous result is unchanged. The cluster members at the turnoff of the color-magnitude diagram are 
used to test and confirm the recently discovered variation in reddening across the face of the cluster, with a probable 
range in the variation of $\Delta$$E(B-V)$ = 0.045 $\pm$ 0.015. With
the slightly higher reddening and lower [Fe/H] compared to commonly adopted values, isochrone fitting leads to an
age of 2.3 $\pm$ 0.2 Gyr for an apparent modulus of $(m-M)$ = 12.40 $\pm$ 0.12.

\end{abstract}


\section{Introduction}
While the mantra of a common distance, age, and composition is invariably cited to justify the study of star clusters as
testbeds of stellar evolution, the reality imposed by cluster dynamical evolution within a Galactic gravitational potential well 
often limits the application of these admittedly valuable characteristics. The typical timescale for evaporation of open clusters is only a 
few $\times$ $10^8$ years, as confirmed by statistical studies of star clusters within a few hundred parsecs of the Sun \citep{LA05}, 
while for clusters interior to the solar circle, the decline in numbers beyond 1 Gyr in age is even more dramatic \citep{JA94, CA14}. 
For clusters at greater distance, contamination of sparsely populated regions of the color-magnitude diagram (CMD) by field stars, 
sometimes complicated by differential reddening, can make delineation of the more rapid phases of post-main-sequence evolution an 
exercise in self-delusion. These limitations heighten the impact of any richly populated, nearby open cluster older than 2 Gyr, as 
exemplified by the prolific literature on iconic clusters such as M67 and NGC 188 and, more recently, NGC 2420 and NGC 6791. A surprisingly
understudied cluster for many years was NGC 6819 \citep{BU71, AU74}, surprising because of its extremely rich, well-defined CMD from 
the main sequence to the tip of the giant branch within the underpopulated cluster age range of 2-4 Gyrs. Its profile has changed 
dramatically in the last few years due to its inclusion within the Kepler field, making it the focus of asteroseismic studies reaching 
down the giant branch \citep{ST11}, with a rapidly expanding literature related to the cluster and its 
members \citep{AT13,JE13,PL13,YA13,WU14,WL14}. 

With the goal of using atmospheric Li to probe stellar structure and evolution among low mass stars, the authors 
have undertaken an extensive spectroscopic program to survey members of a key set of open clusters from the 
tip of the giant branch to as far down the main sequence as the technology allows. First results have been published 
for the clusters NGC 3680 (age = 1.75 Gyr) \citep{AT09} and NGC 6253 (3.0 Gyr) \citep{AT10, CU12}. Among the clusters 
currently under analysis are NGC 7789 (1.5 Gyr) and NGC 6819 (2.3 Gyr), with over 300 stars observed in each. The age of NGC 6819 
places the turnoff stars in a mass range where partial degeneracy at hydrogen exhaustion slows the evolutionary rate enough
to populate both the subgiant branch and the first-ascent giant branch below the red giant clump. Preliminary spectroscopic 
analysis has already led to the discovery of a unique Li-rich giant fainter than the level of the clump \citep{AT13}, below the point 
where standard stellar evolution models predict the initiation of mixing assumed to create Li-rich atmospheres.

However, high dispersion spectroscopic analysis requires reliable input parameters for the models used in interpreting the spectra,
specifically temperatures and surface gravities. These, in turn, demand precise estimates of the cluster reddening
and distance, usually derived from comparison of the observed CMD to theoretical isochrones of appropriate age and metallicity. 
The goal of this paper is to define the fundamental properties of NGC 6819 through an analysis of the cluster on 
the $uvbyCa$H$\beta$ photometric system. The efficacy of this approach has been demonstrated many times in the past, including studies 
of NGC 5822 \citep{CA11}, NGC 6253 \citep{TA03}, NGC 6791 \citep{AT07}, and Mel 71 \citep{TW06}. Of particular value for open clusters 
in rich and extended fields is the photometric capability to derive individual reddening estimates while eliminating probable 
field stars which may have proper motions consistent with cluster membership, a point we will return to in Sec. 3.	

The outline of the paper is as follows: Sec. 2 discusses the CCD observations and their reduction to the 
standard system for intermediate-band photometry; Sec. 3 uses the photometry, in conjunction with proper-motion membership, to
identify and isolate probable cluster members which become the core data set for selecting single, main sequence stars
for reddening and metallicity estimates in Sec. 4. We also test (and confirm) the existence of the reddening gradient
across the face of the cluster as derived by \citet{PL13} (hereinafter referred to as PL). Sec. 5 contains a discussion
of the potential impact of the new reddening and metallicity on the cluster distance and age and a summary of our conclusions.

\section{Observations and Data Reduction}

\subsection{Observations} 
Intermediate and narrow-band imaging of NGC 6819 was completed using the WIYN 0.9-m  telescope on UT dates July 2-7, 2011, 
June 14-20, 2012, and June 8-13, 2013. Fourteen of these nineteen nights were usable for photometry but only a few
were totally photometric. For all nights other than five during  the June 2012 run, the S2KB CCD was used at 
the $f$/7.5 focus of the telescope for a 20\arcmin $\times$ 20\arcmin\ field with 0.6\arcsec\ pixels. $vbyCa$H$\beta$ 
data from five of the nights in 2012 were obtained using the smaller T1KA chip with identical pixel size. The 
psf-based photometry for each smaller frame for each filter was transformed to the instrumental system of the S2KB 
chip using a linear calibration and a color term coupled to $(b-y)$. However, the standard stars observed during 
these five nights have not been utilized for calibration purposes; only standards observed with the S2KB chip have
been used for calibration. All seven filters are from the 3\arcsec $\times$ 3\arcsec\ filter set owned jointly by the University of Kansas and Mt. Laguna Observatory.

Bias frames and dome flats for each filter were obtained every night, with sky flats observed at twilight for the $u$, $v$, and $Ca$
filters when feasible. On all photometric nights, fields in NGC 6819, as well as standard stars and extinction stars, were observed over a range in air mass. Extinction coefficients were separately determined for each photometric night. 

Standard IRAF routines were used to perform initial processing of the frames, i.e. bias-subtraction and flat-fielding.  
Illumination corrections were applied for frames obtained in 2013 and for the shortest wavelength frames obtained in 2012. 
A fairly comprehensive discussion of our procedure for obtaining PSF-based instrumental magnitudes and merging multiple 
frames of a given filter can be found in \citet{AT00}.

Our calibrations to the standard extended Str\"omgren system are based on aperture photometry in the program 
cluster, of field star standards, and of stars in NGC 6633 for each photometric night.  For every frame contributing 
to the photometric calibration solution, aperture magnitudes for standard stars are obtained within apertures scaled 
to five times the FWHM for the frame; sky annuli are uniformly chosen with the inner radius one pixel larger
than the aperture and a uniform annular width. A number of sources were consulted for field star standard index values, 
including the catalog of \citet{TA95} for $V$, $b-y$ and $hk$ indices, catalogs of $uvby$H$\beta$ observations by 
\citet{OL83,OL93,OL94}, and compilations of H$\beta$ indices by \citet{HM98} and \citet{SN89}. 
H$\beta$ indices for stars in NGC 6633, obtained by \citet{ED76}, were used to augment the field star 
standards in calibrating H$\beta$ indices.

Following standard procedures for Str\"omgren photometry, a single $b-y$ calibration equation was derived for
warmer dwarfs and giant stars and a separate calibration equation for dwarfs with $(b-y)_0 \geq 0.42$.
Calibrations of $m_1$ and $c_1$ for cooler giants are determined independently from calibrations applied 
to bluer dwarfs or calibrations applied to cooler main sequence stars. All photoelectric standards, field 
stars and cluster stars alike, were used to determine slopes and color terms for the calibration equations, summarized in Table
1. An independent zero-point was determined for each calibration equation on each night, based on field 
star standards, with the exception that stars in NGC 6633 were used as H$\beta$ standards for the one 
photometric night contributing to the calibration of H$\beta$ indices. 

As is our usual procedure, we extend the calibration equations to the indices for NGC 6819 stars based on merged 
profile-fit photometry 
by determining the average differences between profile-fit indices and indices determined 
from aperture photometry in the cluster on each photometric night. These data presented an unusually challenging 
case for this aperture correction scheme, caused jointly by moderately poor seeing on some nights and a relatively 
crowded cluster core field.  It was necessary to determine the average difference between the aperture and 
profile-fit indices based on carefully selected -- and not very large -- sets of 
stars for which no neighbors would be included inside apertures of $\sim 20$\arcsec\ radius.
With such aperture corrections, the calibration equations from each photometric night may be applied to the aperture photometry 
in the program cluster, and then by extension to the profile-fit indices in NGC 6819, with an independent zero-point determined
for each equation from each photometric night. 

Several indicators of the precision of the calibration equations' zero-points are presented in Table 1. For each photometric 
night, $\sigma_1$ quantifies the dispersion of calibrated values about standard values for field star standards; $\sigma_2$ 
quantifies the dispersion among zero-points for the several photometric nights.  Both $\sigma$ values are standard deviations. 
The final calibration equation's zero-point is determined from a weighted sum of the independent night evaluations; the 
final statistic labeled ``sem" denotes the standard error of the weighted mean. 

Final photometry on the $uvbyCa$H$\beta$ system can be found in Table 2, where most columns are self-explanatory:
(X,Y) CCD positions have been translated to the right ascension and declination coordinate frame of PL. As described further in Section 3, we were able to match
a majority of our photometric sample to the positions and photometry prsented in PL; membership probabilities for the nearly 4000 matched stars are included in Table 2 where available.  Stars are 
included in Table 2 only if they were observed at least twice in both $y$ and $b$ filters to construct $V$ and $b-y$, 
twice each in $v$, $Ca$, and $u$ for $m_1$, $hk$, and $c_1$, respectively. Two observations each in the narrow and wide 
filter were required for the H$\beta$ index to be retained. Standard errors of the mean for the indices are calculated by combining the errors for 
individual filters in quadrature and are defined solely by the internal, i.e. frame-to-frame, precision of the of the individual filters. The increase in scatter among the indices for the brightest stars is due the inclusion of two bright, very red giants that exhibit apparent variability and to the reduced  number of CCD frames with exposure times short enough to leave the brightest stars unsaturated.  A plot of the average {\it sem} for each index as a function of 
$V$ can be seen in Fig. 1.  The longer, on-line version of Table 2 includes
photometry for 7187 stars, subject to the limitations described above as well as restriction to stars with $\sigma_{b-y} \leq 0.10$.

\subsection{Comparison to Previous Photometry} 
Our $V$ magnitudes can be compared directly to 
the three comprehensive broad-band surveys to date which include $V$ photometry \citep{RV98, KA01, YA13}
(hereinafter referred to as RV, KA, and YA, respectively) with some surprising results. Beginning with RV, 
in the bottom panel of Fig. 2 we show the residuals in $V$, in the sense (RV - Table 2), for 1690 stars matched via the WEBDA (X,Y) 
coordinates for RV and our data. No known variables or stars with larger than average errors have been eliminated 
from the figure. The overall pattern exhibits a mean offset close to 0.00 at all magnitudes with increasing 
scatter as $V$ approaches 18. Somewhat surprising is the dispersion in the residuals among the brighter stars. 
One would expect that, given the high precision of both photometric surveys at $V$ = 16 and brighter, the trend 
should narrow significantly toward smaller $V$. 

After running a number of tests, the source of the scatter emerged as a gradient in the $V$ photometry 
with right ascension or X in WEBDA coordinates. Using 566 stars brighter than $V$ = 16 to minimize the impact 
of larger photometric scatter, we plot the residuals in $V$ as a function of X CCD position in the top panel of Fig. 2. 
The trend is obvious and is confirmed even if stars at all magnitudes are included. Fitting a linear relation to the data, the residual gradient becomes

\noindent
$\Delta V = (0.003 \pm 0.001) - (0.043 \pm 0.002) \times X (kilopixels)$
 
\noindent

Applying this to the photometry of RV, the revised residual plot is shown in the middle panel of Figure 2. The improvement is obvious; the mean 
residual is, by definition, 0.000 with a dispersion of only $\pm$0.016 for $V\leq 16$ and $\pm$0.029 for all stars in the sample. 
Again, using only stars brighter than $V$ = 16, no correlations of significance were found for the residuals with either 
$B-V$ color or declination (Y). We can attribute the trend to the photometry of RV because a comparison 
between KA and RV shows the same gradient. We also note that position-dependent gradients of similar size have 
been identified in the broad-band data of \citet{GI98} for NGC 7789 \citep{TW13}.

Turning next to KA, we plot in Fig. 3 the residuals in $V$, in the sense (KA - Table 2), for 1819 stars 
common to the two surveys. We emphasize that the residuals have a slightly larger range than in Fig. 2 and no stars 
with residuals more negative than -0.10 mag have been excluded from the plot. Two trends are obvious: 

(1) There is a distinct dichotomy among the stars above and below $V$ = 14. The mean offset among the brighter sample 
is about -0.01 mag but the scatter is excessive, while the fainter sample shows a mean offset closer to 0.02 mag. 

(2) Among the fainter stars, the dispersion is highly asymmetric, with a long and well-populated tail toward larger 
residuals, virtually independent of the magnitude.

Both of these trends are confirmed through comparisons between RV and KA, with or without a spatial correction to the 
data of RV. Among the brighter sample, there is weak evidence for a spatial dichotomy in the photometric zero-point coupled
to declination, with stars in the north being offset from those in the southern half of the field by 0.03 to 0.04 mag, but the
sample is small. For the stars fainter than $V$ = 14, no obvious spatial gradient emerges which would explain the extended tail toward
positive residuals. The asymmetry may be a reflection of the way in which crowded stellar images have been handled by the PSF 
software. KA provided no comparisons with the photometry of RV so this issue was never addressed.

The third and final large survey is that of YA in $VI$. While YA include some discussion of the residuals between their
work and earlier studies, noting good agreement, their comparison sample includes fewer than 100 stars, more than an order of
magnitude fewer than we can readily find in common with RV and KA. The reason for this deficiency remains a mystery.
For comparison purposes, the data of Table 2 were cross-matched with those of KA after the CCD coordinates for both samples were
transformed to right ascension and declination, as defined in PL. Using a match radius of 1\arcsec\ and eliminating any
stars with magnitude differences greater than 0.2 mag to minimize mismatches leaves a common sample of 1619 stars to $V$ = 18.5. 
The residuals in $V$, in the sense (YA - Table 2), are plotted in Figure 4. The agreement with YA is excellent, with a modest 
asymmetry to the positive side of the residuals. From 516 stars brighter than $V$ = 16.0, the mean offset is +0.003 $\pm$ 0.023; 
for all 1619 stars, the mean residual is +0.004 $\pm$ 0.028, where the quoted errors are standard deviations.

We conclude that our $V$ photometry is on the same system as YA and RV, corrected for a spatial gradient, to better 
than a few millimagnitudes in the mean. By contrast, the sample of KA exhibits a distinct change in zero point at 
$V$ $\sim$ 14, with the fainter sample approximately 0.03 mag too faint relative to the other three studies.

\section{The Color-Magnitude Diagram}
The CMD based upon ($V, b-y$) for all stars in Table 2 is shown in Fig. 5. 
Stars with {\it sem} errors in $b-y$ below 0.015 are shown as open circles, while stars with errors greater than 
this cutoff to an {\it sem} limit of 0.15 mag are plotted as crosses. While some primary features of the CMD, the location of the
turnoff and the red giant clump, are obvious, due to the area of the sky covered by the frames, field star confusion makes
delineation of the giant branch below the clump and the fainter main sequence almost impossible. 

As a first attempt to minimize the field star contamination, we plot in Fig. 6 all stars within 5\arcmin\ of the cluster center as defined
by PL. The improvement in isolating the cluster is readily apparent, though the reduction in both field stars and
cluster members for $V \geq 18$ still leaves a respectable level of contamination. The cluster core exhibits a rich population of blue
stragglers, a well-defined red clump and bright giant branch, but a non-negligible degree of confusion among potential subgiants and
first-ascent red giants below the clump.

Fortunately, the astrometric analysis of the cluster by PL supplies proper-motion membership for the majority of the stars in our field.
As stated in Sec. 2, we have transformed our CCD (X,Y) coordinates to the (RA,DEC) system of PL, which should be J2000, epoch 2009.875.
As a first step, all stars' coordinates common to both data sets and coincident within a radial distance of 2\arcsec\ were identified. 
Next, the stars common to YA and PL were used to derive a transformation between the $g$, $g-r$ photometry of PL and the $V, V-I$ data of YA.
From 1250 stars brighter than $V$ = 18, we found
     
\noindent
$ V = g -(0.038 \pm 0.006) - (0.595 \pm 0.018)(g-r) + (0.061 \pm 0.012)(g-r)^2 $

\noindent
$V-I = (0.403 \pm 0.010) - (0.408 \pm 0.029)(g-r) + (0.358 \pm 0.020)(g-r)^2 $

\noindent

The dispersions among the residuals in $\Delta$$V$ and $\Delta$$(V-I)$ are 0.022 mag and 0.029 mag, respectively. With the transformed 
$g$ photometry in hand, any star common to PL and Table 2 through RA and DEC which showed a difference in $V$ greater than 0.1 mag was 
excluded from the sample. Finally, all stars with membership probabilities less than 50\% were excluded. Three stars brighter than 
$V$ = 17 with proper-motion probabilities above 50\% which had been excluded due to an excessive $V$ residual were reinstated. All 
other excluded stars from the original match are either fainter than $V$ = 17 and/or have membership probabilities below 50\%. The 
resulting ($V, b-y$) CMD for all members within the CCD frames is given in Fig. 7. Symbols
have the same meaning as before, but the sample has been cut at $V$ = 19 to reduce scatter caused by increasing errors in $b-y$.

The improvement relative to even Fig. 6 is dramatic. While all the primary features from the core sample CMD remain, a plausible outline
of the subgiant branch and first-ascent giant branch below the clump now emerges. The main sequence is tightly defined, with only a modest
level of scatter to the red, as expected for the band defined by probable binaries extending up to 0.75 mag above the unevolved main sequence.
The unevolved main sequence can now be traced down to the limit of the plot, though it is apparent that some of the background field stars
with proper motions comparable to the cluster still remain near $b-y$ = 0.5, as well as their evolved counterparts in the vertical band between
$b-y$ = 0.7 and 0.8.

\section{Cluster Properties - Reddening and Metallicity}
With the restricted sample of Fig. 7, we can now approach the photometric reddening and metallicity estimate. Reddening on the $uvby$H$\beta$
system is defined by comparison of the predicted $b-y$ color tied to the stellar temperature as derived from the reddening-free H$\beta$
index, adjusted for evolutionary effects and metallicity, to the observed $b-y$ index for each F and early G cluster dwarf. Potential
sources of scatter in the final cluster averages include poor photometry, non-members, composite systems with distorted photometry, and
inadequate correction for the effects of post-main-sequence evolution. As we have done consistently in the past, our approach is to eliminate
any stars which, for any of the reasons noted above, might reduce the reliability or skew the estimate of the final cluster parameters. 
The first simple cut is to eliminate all stars brighter than $V$ = 15.75, i.e. stars significantly evolved beyond the main sequence 
and/or populating the turnoff in a region likely to be contaminated by the extended binary sequence, and all stars fainter 
than $V$ = 17.50, where photometric errors in indices other than $b-y$ begin to grow rapidly with increasing magnitude. An 
expanded ($V$, $b-y$) CMD for this sample is shown in Fig. 8.

The delineation of a tight main sequence with a full range of only 0.03 to 0.04 mags in $b-y$ at a given $V$ is encouraging, though some 
stars do lie well off the main sequence. To isolate likely single {\it member} stars, we have drawn a mean linear relation through the main sequence and
tagged any star within 0.02 mag at each $V$ as a probable single-star member (open circles). Stars 0.02 to 0.07 mag redder than the mean relation  
are defined as probable binaries (crosses); stars at least 0.02 bluer than or $\geq 0.07$ mag redder than the main line 
are tagged as probable non-members (filled triangles). This leads to a sample of 382 probable single-star (or binaries with low mass ratios) members.

An immediate question in light of claims for potential reddening variation across the face of the cluster (PL) is whether or not the red
limit eliminates true cluster members with higher than average reddening rather than binaries or non-members. Fortunately, $hk$ photometry 
offers an effective resolution. The $hk$ index is designed to supply a metallicity estimate for stars of a given temperature but,
for stars of a given metallicity, it is a strong function of temperature. However, unlike $b-y$, $hk$ is only weakly dependent on reddening,
with $E(hk)$ = -0.16$E(b-y)$. Thus a star which appears too red at a given $V$ in Fig. 8 because it is intrinsically cooler will shift in a
($V, hk$) diagram by an even larger amount. If the star is shifted in $b-y$ due to enhanced reddening, the shift in ($V, hk$) will be
small to negligible and toward the blue, i.e. smaller $hk$. The ($V, hk$) diagram is shown in Fig. 9; symbols have the same meaning as in Fig. 8. 
All stars with errors in $hk$ larger than 0.03 mags have been eliminated, reducing the sample by 17 stars, including four stars 
classed as singles in Fig. 9. While some stars classed as deviants in Fig. 8 do lie on the main sequence in Fig. 9, the 
classifications for the majority are confirmed. Additionally, a half dozen stars below $V$ = 16.9 appear significantly bluer 
than expected (filled circles). Based upon H$\beta$ and $m_1$ indices, these are 
either metal-deficient background stars or heavily reddened, hotter dwarfs; neither group contains probable cluster members. 
Based upon the positions in Fig. 9, we eliminate 12 additional stars which may be binaries or field stars, leaving a sample of 366 stars. 

To ensure that only the most precise photometry is included in the analysis, we eliminate all stars with errors in H$\beta$, $m_1$, 
and $hk$ larger than 0.02 mag, 0.02 mag, and 0.03 mag, respectively. Finally, color transformations between H$\beta$ and $b-y$ 
invariably include a correction for evolution, defined via the $c_1$ index, with more evolved stars having larger $c_1$ values than 
an unevolved star on the main sequence. While the majority of stars in our restricted sample have measured $c_1$ indices, since we 
are extending the data below $V$ = 16.5, the errors in $c_1$ grow rapidly with increasing $V$, as indicated by Fig. 1. To avoid 
elimination of a large fraction of the data set, we have chosen to adopt the $c_1$ value at the observed H$\beta$ as defined by 
the standard relation as the reddening-corrected $c_1$ index for each star. In short, we have assumed that all stars are 
unevolved, in keeping with the restriction imposed by eliminating all stars brighter than $V$ = 15.75. 

As in past cluster analyses, use is made of two intrinsic H$\beta$-$(b-y)_0$ relations to define the intrinsic colors. The intrinsic $b-y$
color is derived in iterative fashion, starting with an assumed $E(b-y)$ and metallicity, calculating the cluster metallicity, redefining the
intrinsic color for the calculated metallicity to obtain a new reddening, and repeating the sequence until the change in reddening
is too small to be statistically significant.  The first $E(b-y)$ from \citet{OL88} applies to F stars in the H$\beta$ range from 
2.58 to 2.72. From 278 F dwarfs that meet all the criteria, the mean $E(b-y)$ is found to be 0.115 $\pm$ 0.016 (sd). The second 
intrinsic color relation is that of \citet{NI88}, a slightly modified version of the original relations derived by \citet{CR75, CR79} 
for F and A stars. For the same 278 F dwarfs, the alternate relation implies $E(b-y)$ = 0.119 $\pm$ 0.017 (sd). It should be noted that the 
slightly higher reddening for F stars using the \citet{NI88} relation compared to that of \citet{OL88} is a consistent 
occurrence from such comparisons \citep{TW06, AT07}. A weighted average of the results leads to $E(b-y)$ = 0.117 $\pm$ 0.002 (sem), or 
$E(B-V)$ = 0.160 $\pm$ 0.003 (sem). When combined with the zero-point uncertainties in $b-y$ and H$\beta$, the total uncertainties 
in $E(b-y)$ and $E(B-V)$ become $\pm$0.005 mag and $\pm$0.007 mag, respectively. It should be emphasized that the dominant source of 
uncertainty in the reddening is the zero-point of the $b-y$ photometry since the $(b-y)$, H$\beta$ relation has a relatively shallow 
slope over the color range of interest, as seen in Fig. 11 of \citet{CA11}.

With the reddening fixed, the next step is the derivation of metallicity, a parameter that can be defined using $hk$ or
$m_1$ coupled to either $b-y$ or H$\beta$ as the primary temperature indicator. In past studies using $uvbyCa$H$\beta$ photometry,
the metallicity from $hk$ tied to H$\beta$ invariably has been given the greatest weight due to the greater sensitivity of $hk$ to
modest metallicity changes, while the H$\beta$-based relations allow decoupling between errors in the two indices and minimize the impact
of potential reddening variations, if any exist. We will follow the same approach with NGC 6819, allowing us to tie our results 
directly into the same metallicity scale from past intermediate-band cluster studies. 

With $E(b-y)$ = 0.117, the mean $\delta$$m_1$($\beta$) for 282 F dwarf probable members between H$\beta$ = 2.58 and 2.72 
is 0.025 $\pm$ 0.001 (sem), where $\delta$$m_1$ = 0.0 is set at the adopted Hyades metallicity of +0.12. On this same scale, 
NGC 3680, NGC 5822 and IC 4651 respectively have $\delta$$m_1$ = 0.027 $\pm$ 0.002 (sem) \citep{AT04},
+0.017 $\pm$ 0.003 (sem) \citep{CA11}, and 0.000 $\pm$ 0.002 (sem) \citep{AT00}, implying that NGC 6819 is clearly lower in [Fe/H] 
than the Hyades, and almost as deficient as NGC 3680, a somewhat surprising result given the consistent claim that the
cluster is metal-rich, a point we will return to in Sec. 4.2. The $\delta$$m_1$ measure translates to [Fe/H] = -0.116 $\pm$ 0.012 (sem) 
on a scale where NGC 3680, NGC 5822, and IC 4651 have [Fe/H] = -0.17, -0.06, and +0.12, respectively. The translation from
index to metallicity is partially dependent upon the mean color/temperature of the sample, which explains the modest
shift in the relative ranking of the clusters in switching from index to [Fe/H]. As an additional reference point, 
the photoelectric $uvby$H$\beta$ data of M67 produce [Fe/H] = -0.06 $\pm$ 0.07 \citep{NI87}. Taking into account the uncertainty
in the zero-point of the $m_1$ indices, [Fe/H] = -0.12 $\pm$ 0.10 with internal and external errors combined.

Turning to the $hk$ index for 282 stars, $\delta$$hk$($\beta$) = 0.050 $\pm$ 0.003 (sem), which translates to 
[Fe/H] = -0.055 $\pm$ 0.010 (sem), on a scale where [Fe/H] = +0.12, 0.01, and -0.10 for the Hyades, NGC 5822, and NGC 3680. 
Taking errors in the $hk$ zero-points into account, [Fe/H] = -0.06 $\pm$ 0.03, internal and external uncertainty combined.
A weighted average of the two metallicity estimates leads to [Fe/H] = -0.06 $\pm$ 0.04, where the errors refer to the 
combined internal and external errors from the combined indices.

Before discussing the significance of these results, an obvious question is the impact of variable reddening on our conclusions. For the
[Fe/H] derived from $hk$, the impact is negligible. By definition, H$\beta$ is reddening-independent. If we include a range 
of 0.08 in $E(V-I)$ among the stars in the cluster, this translates to a range of 0.007 in $E(hk)$, or a spread of $\sim$0.025 in [Fe/H]. 
For $m_1$, the  greater impact of reddening on the index coupled to a steeper $\delta$$m_1$ - [Fe/H] relation translates the same reddening range to
an [Fe/H] range seven times larger. Fortunately, we can derive the reddening for each star individually, rather than
adopting the cluster mean for all stars, and recalculate the cluster metallicity . If we adopt the photometric reddening estimate from $b-y$, H$\beta$
for each star, the resulting mean [Fe/H] from $m_1$ becomes -0.120
$\pm$ 0.013 while the analogous estimate from $hk$ is [Fe/H] = -0.055 $\pm$ 0.011 (sem), leading to a weighted average [Fe/H] = -0.06 $\pm$ 0.04.
Given the very modest range derived below for $E(b-y)$ and the impact of even small photometric scatter, the lack of a statistically 
significant change in the mean abundances is expected.

\subsection{Reddening Variability}
As discussed in Sec. 1, one of the potential sources of parametric scatter for a cluster spread over a field 20\arcmin\  across at a distance of
more than a kiloparsec is variable reddening along the line of sight. Due to the well-defined nature of the CMD at the turnoff plotted
in Fig. 7, we can immediately use the full width of the turnoff in $b-y$ at a fixed $V$ to place 
an upper limit of 0.045 mag on the range in $E(b-y)$, 0.060 for $E(B-V)$, without applying any
adjustment for photometric scatter in $b-y$ and/or contamination by a binary sequence. PL have used the $VI$ data of YA for proper-motion
members to define a blue edge in the ($V, VI$) CMD at the cluster turnoff, the color region where differential reddening effects should
be maximized due to the combined shift in $V$ and $V-I$. The color offset for each star relative to this fiducial relation is adopted
as an estimate of the differential reddening, up to a limit of $\delta$$E(V-I)$ = 0.09, where binaries may begin to dominate. The 
positional averages of these values are then used to construct a spatial reddening map (see Fig. 10 of PL), showing that 
stars $\geq 2$\arcmin\ east of the cluster center are typically 0.05 to 0.07 mag redder in $E(B-V)$ than the low 
reddening region in the southwest quadrant of the cluster, i.e. for stars $\geq 2$\arcmin\ west and $\geq$ 2 \arcmin\ south of the cluster center.

To test this claim, we have used the photometry defining Fig. 7 and isolated two distinct samples. The group expected to be more highly 
reddened is composed of all stars $\geq 2$\arcmin\ east of the cluster center while the predicted blue group encompasses stars in the southwest 
quadrant, $\geq$ 2\arcmin\ west and south of the cluster center. The CMD for the expanded region near the turnoff is shown in Fig. 10a with
the colors and symbol types indicating which reddening group the star belongs to (blue triangles or red crosses). 
The pattern is obvious; there is little doubt that the reddening trend 
established by PL is real. We can place a tight constraint on the size of the range between these two regions by shifting the blue stars 
appropriately in $b-y$ and $V$ until the CMDs overlap. Fig. 10b shows the impact of adding 0.022 (0.030) mag of reddening in $E(b-y)$ ($E(B-V)$) 
to the blue stars of the southwest quadrant. The scatter in the turnoff
is cut in half, reduced almost to what one would expect from photometric scatter alone. Since the mean shift in color between the two 
regions should be less than the most extreme variation across the cluster, we estimate that the true spread in $E(B-V)$ lies somewhere 
between the 0.03 estimate above and the point-to-point range of 0.06 found by PL or $\Delta$$E(B-V)$ = 0.045 $\pm$ 0.015.

\subsection{Previous Reddening and Metallicity Estimates}
With the confirmation that NGC 6819 suffers from variable reddening in a range of 0.03 to 0.06 mag in $E(B-V)$, the import of past 
reddening estimates is reduced since the values derived will depend in part on where in the cluster the reddening was measured.
A partial summary of derived and adopted reddening estimates for NGC 6819 can be found in \citet{WL14}. With the exception of two early studies
by \citet{AU74} and \citet{LI72}, and those studies which later adopted these flawed estimates \citep{FJ93, TW97}, the range among more
recent analyses is $E(B-V)$ = 0.10 to 0.16. Care must be taken, however, in assigning weight to many of these values. For
example, \citet{KA01} adopts without explanation $E(B-V)$ = 0.10, a value later assumed without explanation by \citet{HO09}. 
\citet{AT13} use the mean of the derived literature values between 0.12 and 0.16. \citet{JE13} attempt to
derive the cluster reddening through a differential comparison of the clump stars in M67 and NGC 6819. However, the final value is
built upon the assumption that the clump stars in NGC 6819 have uniform reddening and that NGC 6819 is more metal-rich than M67 by
0.05 dex. \citet{BA13} investigate map-based $A_V$ estimates from the Kepler Input Catalog \citep{BR11} to obtain $E(B-V)$ = 0.189 $\pm$ 0.002 
for 564 stars within 12\arcmin\ of the cluster; the mean reddening rises to 0.203 $\pm$ 0.003 for cluster members. However, they
default to 0.15, referring back to RV and the spectroscopic results of \citet{BR01}. \citet{WU14} derive reddening through an isochrone
match to the $BV$ data of \citet{HO09}, arriving at a simultaneous estimate of the reddening, metallicity, and distance with $E(B-V)$
= 0.13 and an unrealistically small uncertainty of $\pm$0.01. In the most recent paper, \citet{WL14} simply default to the mean supplied by
\citet{BR01}. In summary, there are few direct, reliable estimates of the reddening to NGC 6819 which don't presume either uniform
reddening and/or an assumed high metallicity. Only RV derived $E(B-V)$ = 0.16 and [Fe/H] $\sim$-0.05 through isochrone matches to the $BV$ CMD.
\citet{BR01} represents the most cited source for a reddening ($E(B-V)$ = 0.14 $\pm$ 0.04) and metallicity 
estimate ([Fe/H] = +0.09 $\pm$ 0.03), but these parameters are defined by 3 red clump giants, one of which is now known to 
be a binary \citep{HO09}. \citet{FJ93} derived [Fe/H] = +0.05 from moderate-dispersion spectroscopy, adopting $E(B-V)$ = 0.28 on a scale
where M67 had [Fe/H] = -0.09. The differential was revised by \citet{TW97} using the same excessive reddening value, finding [Fe/H] =
0.07 on a scale where M67 has [Fe/H] = 0.00. If one adopts the correction to [Fe/H] of -0.06 dex for a decline of 0.05 in $E(B-V)$ \citep{FJ93}, 
a shift from $E(B-V)$ = 0.28 to 0.16 should lower the metallicity of NGC 6819 by 0.14 dex, placing it between the metallicity
of M67 and -0.07 dex lower. However, \citet{FR02} revise the reddening of NGC 6819 to $E(B-V)$ = 0.16 and find [Fe/H] -0.11 for the cluster
on a scale where M67 is fixed at -0.15. We conclude that NGC 6819 is most probably more metal poor than M67 by about -0.05 dex, rather
than comparable to the Hyades in metallicity.

\section{Summary and Conclusions}
Recently the old open cluster, NGC 6819, has become a high profile object of considerable astrophysical interest due to its location within
the Kepler field. However, apart from this particular circumstance, the cluster still would be invaluable to those focused
on stellar and galactic evolution because of its rich stellar population and an age between 2 and 4 Gyr, a range shared by very few  
nearby objects. With the goal of deriving high-dispersion spectroscopic abundances for an extensive 
sample of over 300 stars, an intermediate-band photometric program was undertaken to derive the key cluster parameters of
reddening and metallicity as input for estimating individual stellar parameters tied to the cluster age and distance. The need for
such a study is heightened when the literature parameters for the cluster are reviewed; direct reddening estimates 
are few in number, low in quality, and/or often dependent upon an assumed high metallicity tied primarily to spectroscopy of 3 red clump
stars \citep{BR01}. Equally important, greater photometric and astrometric scrutiny of the cluster has led to a claim that the
cluster suffers from variable reddening (PL), an effect never included in past derivations of the reddening or metallicity.

Using $uvbyCa$H$\beta$ photometry of a 20\arcmin\ by 20\arcmin\ field of NGC 6819 collected over three years, high precision photometric indices
have been combined with proper-motion membership to isolate a sample of 382 single-star, unevolved F-G dwarfs on the cluster main
sequence. Restricting this sample further to only 278 stars with high precision measures in all indices within the prescribed limits of the photometric
calibrations leads to a mean reddening of $E(b-y)$ = 0.117 $\pm$ 0.005 or $E(B-V)$ = 0.160 $\pm$ 0.007, internal and external uncertainties
included. Somewhat surprising, even with the slightly higher than usual reddening, the cluster metallicity, when derived using individual
reddening values for each star or adopting a mean value for all stars, becomes [Fe/H] = -0.06, well below what has become the
canonical value of +0.09. Equally important, the exceptional precision of the ($V, b-y$) photometry supplies a convincing confirmation
that the cluster has higher reddening on the east side and lower reddening in the southwestern quadrant of the field. The minimum range 
in variation that we derive as a lower limit (0.03 mag in $E(B-V)$) is somewhat smaller than the maximum range found by PL (0.07 mag), but
this may depend upon the spatial resolution of the mapping and the ability to eliminate the effect of binaries from the main sequence
broadening. We conclude that the  range probably lies between 0.03 and 0.06 mag in $E(B-V)$. 

We close by estimating the broad impact of these changes on the derivation of the cluster distance and age, leaving a 
detailed star-by-star correction for a future paper dealing with the individual spectroscopic abundances \citep{LB14}.
Recent estimates of the cluster distance and age via a variety of techniques typically fall between $(m-M)$ = 12.25 and 12.50 and
1.9 to 2.6 Gyrs, respectively \citep{WL14}. With the exception of RV, the studies regularly used $E(B-V)$ below 0.16 and solar metallicity
or higher, with heavy emphasis on the metal-rich value from \citet{BR01}.

To test the revised reddening and metallicity, we use the $BV$ photometry of RV, cross-identified with PL and selected to only include
members with probabilities above 50\%. We have applied a correction in $V$ to the photometry of RV to adjust for the position-dependent 
offset illustrated in Fig. 2. As a quick correction to the variable reddening, we have adjusted all the colors in the eastern
zone discussed in Sec. 4 by -0.015 mag in $B-V$, all the colors in the delineated southwestern zone by +0.015, and left all other stars
unchanged. The resulting CMD is shown in Fig. 11; the main sequence and the giant branch are surprisingly tight given the simple 
approach to adjusting the reddening, a confirmation of the small range in reddening across the cluster face.  For comparison, we have
superposed a set of $Y^2$ \citep{DE04} isochrones with ages 2.3 (blue, solid line) and 2.5 (red, dashed line) Gyr on the data. The metallicity adopted is [Fe/H] = -0.06, 
with $E(B-V)$ = 0.16. The zero-points of the isochrones have been adjusted by -0.03 and +0.008 in $M_V$ and $B-V$, respectively, to
place them on the same system as our consistently adopted solar values.The fit to the data is very good to excellent for an age of 2.3 $\pm$ 0.2
Gyr. The distance modulus is slightly smaller than that found in \citet{AT13} due to the competing effects of raising the reddening, which should
increase the apparent modulus by 0.11 mag, and lowering the metallicity by 0.15 dex, which will decrease the distance modulus by 0.18 mag \citep{TW09}.
Taking into account the estimated uncertainties in the absolute reddening (0.007 mag), the metallicity scale (0.05 dex), and a conservative estimate
of $\pm$0.10 mag in the ideal vertical fit of the isochrones to the data, all else being equal, one arrives at $(m-M)$ = 12.40 $\pm$ 0.12.
It should be noted that this result is virtually identical to the conclusions reached by RV using a different set of isochrones; the slightly
larger age (2.4 Gyr) and smaller distance modulus (12.35) are completely attributable to a bluer adopted solar color to zero their isochrones.

\acknowledgments
Extensive use was made of the WEBDA database maintained by E. Paunzen at the University of Vienna, Austria (http://www.univie.ac.at/webda). 
The filters used in the program were obtained by BJAT and BAT through NSF grant AST-0321247 to the University of Kansas. NSF support for 
this project was provided to BJAT and BAT through NSF grant AST-1211621, and to CPD through NSF grant AST-1211699. Observing support by
undergraduate Brian Schafer is gratefully acknowledged.

\clearpage

\begin{figure}
\includegraphics[width=\columnwidth,angle=270,scale=0.80]{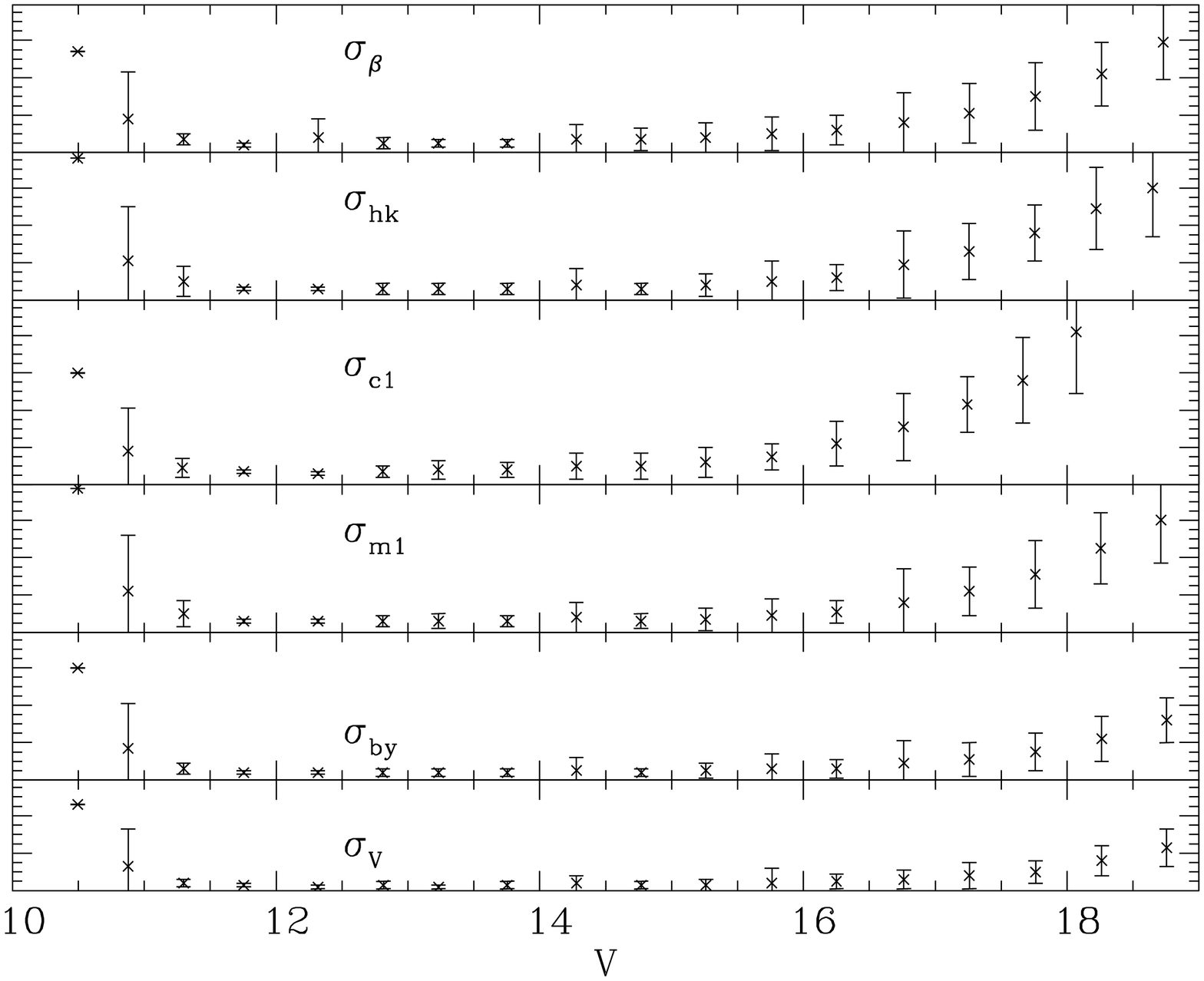}
\caption{Average photometric errors in $V$, $(b-y)$, $m_1$, $c_1$, $hk$ and $H\beta$ as a function of the $V$ magnitude. 
The panel heights are scaled proportionately to the physical range, with major tick-marks indicative of 0.02 mag.}
\end{figure}
\clearpage

\begin{figure}
\includegraphics[width=\columnwidth,angle=270,scale=0.80]{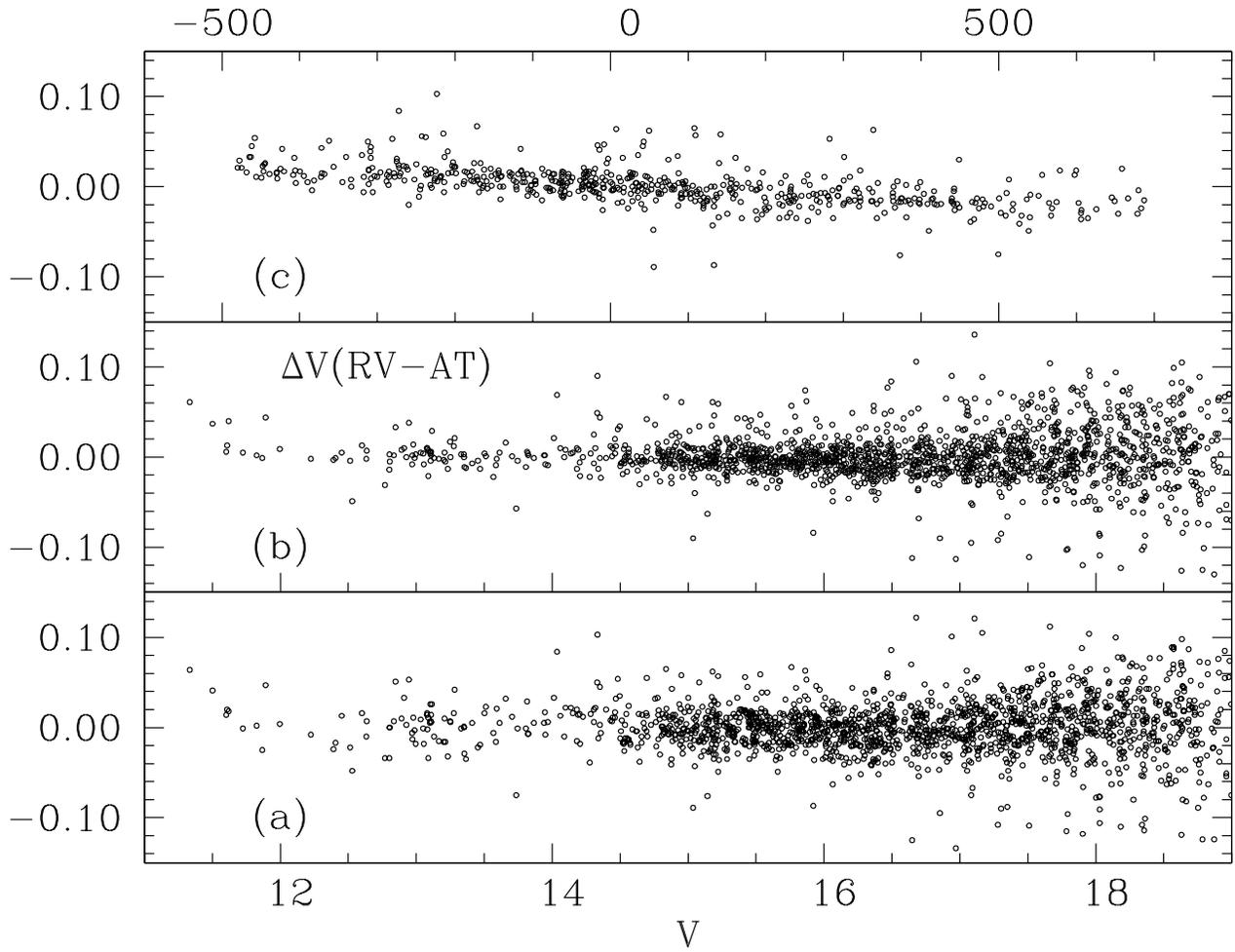}
\caption{(a) Residuals in $\Delta$$V$ as a function of $V$ for (RV-Table 1); (b) Same as (a) after correction for the trend in 
(c), which shows residuals in $\Delta$$V$ as a function of WEBDA coordinate X. }
\end{figure}

\clearpage
\begin{figure}
\includegraphics[width=\columnwidth,angle=270, scale=0.80]{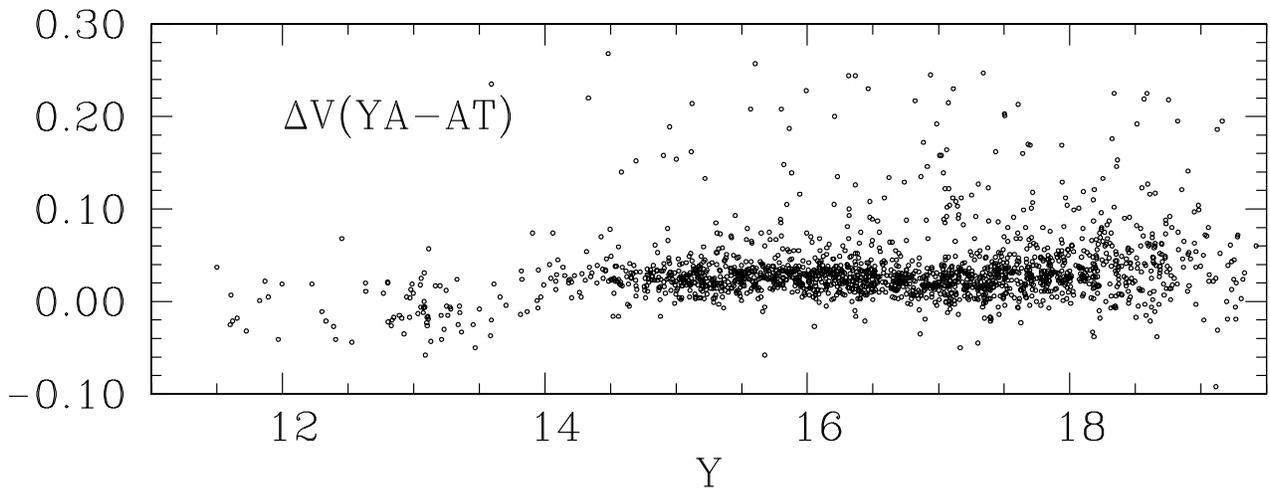}
\caption{Residuals in $\Delta$$V$ as a function of $V$ in the sense (KA - Table 1). The physical scale matches that of Fig. 2.}
\end{figure}

\begin{figure}
\includegraphics[width=\columnwidth,angle=270, scale=0.80]{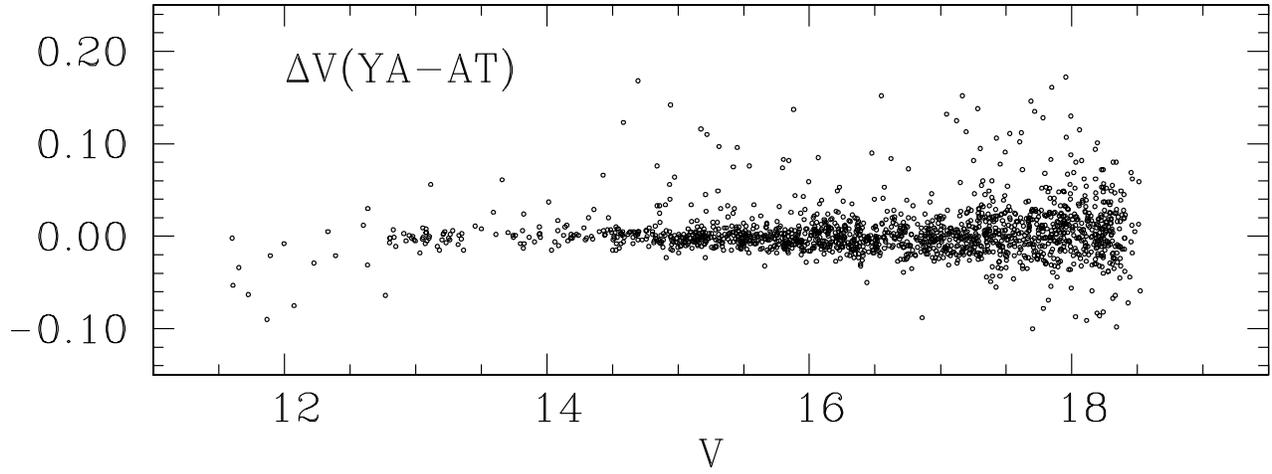}
\caption{Residuals in $\Delta$$V$ as a function of $V$ in the sense (YA - Table 1). The physical scale matches Figures 2 and 3.}
\end{figure}

\clearpage
\begin{figure}
\includegraphics[width=\columnwidth,angle=270, scale=0.80]{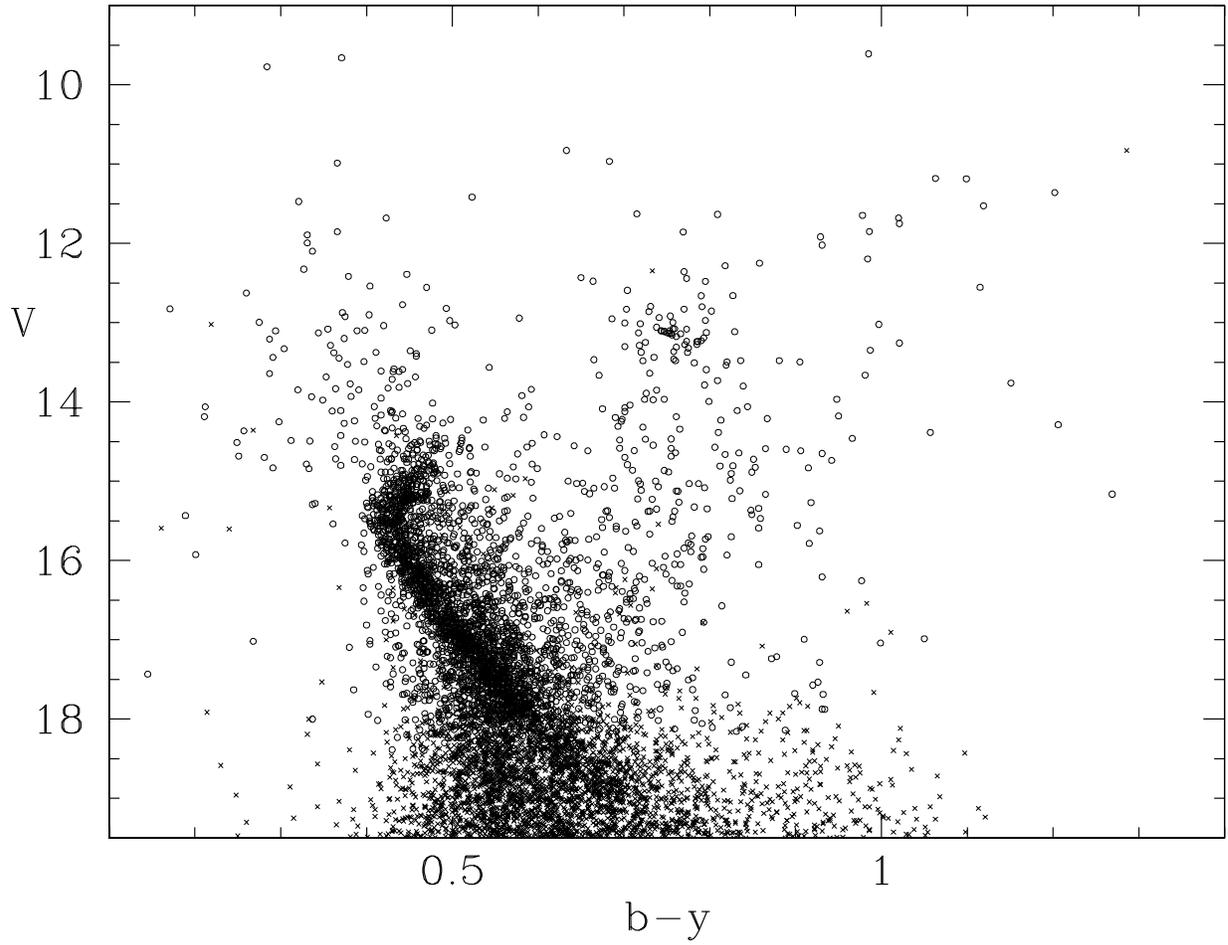}
\caption{CMD for all stars in Table 1. Open circles are stars with photometric errors in $b-y$ $\leq$ 0.015. Crosses are
stars with errors in $b-y$ larger than 0.015. }
\end{figure}

\clearpage
\begin{figure}
\includegraphics[width=\columnwidth,angle=270, scale=0.80]{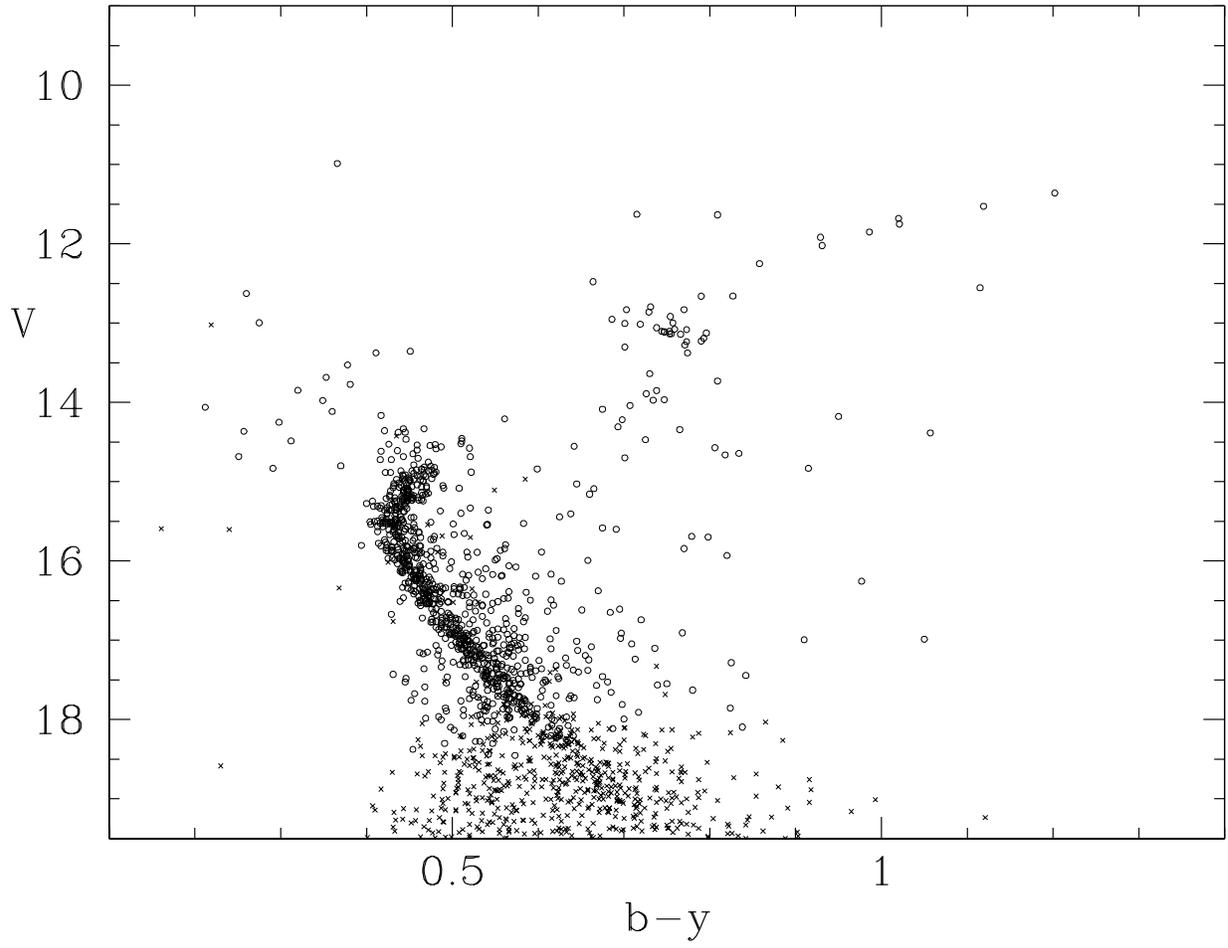}
\caption{Same as Fig. 5 for all stars within 5\arcmin\ of the cluster core.}
\end{figure}

\clearpage
\begin{figure}
\includegraphics[width=\columnwidth,angle=270, scale=0.80]{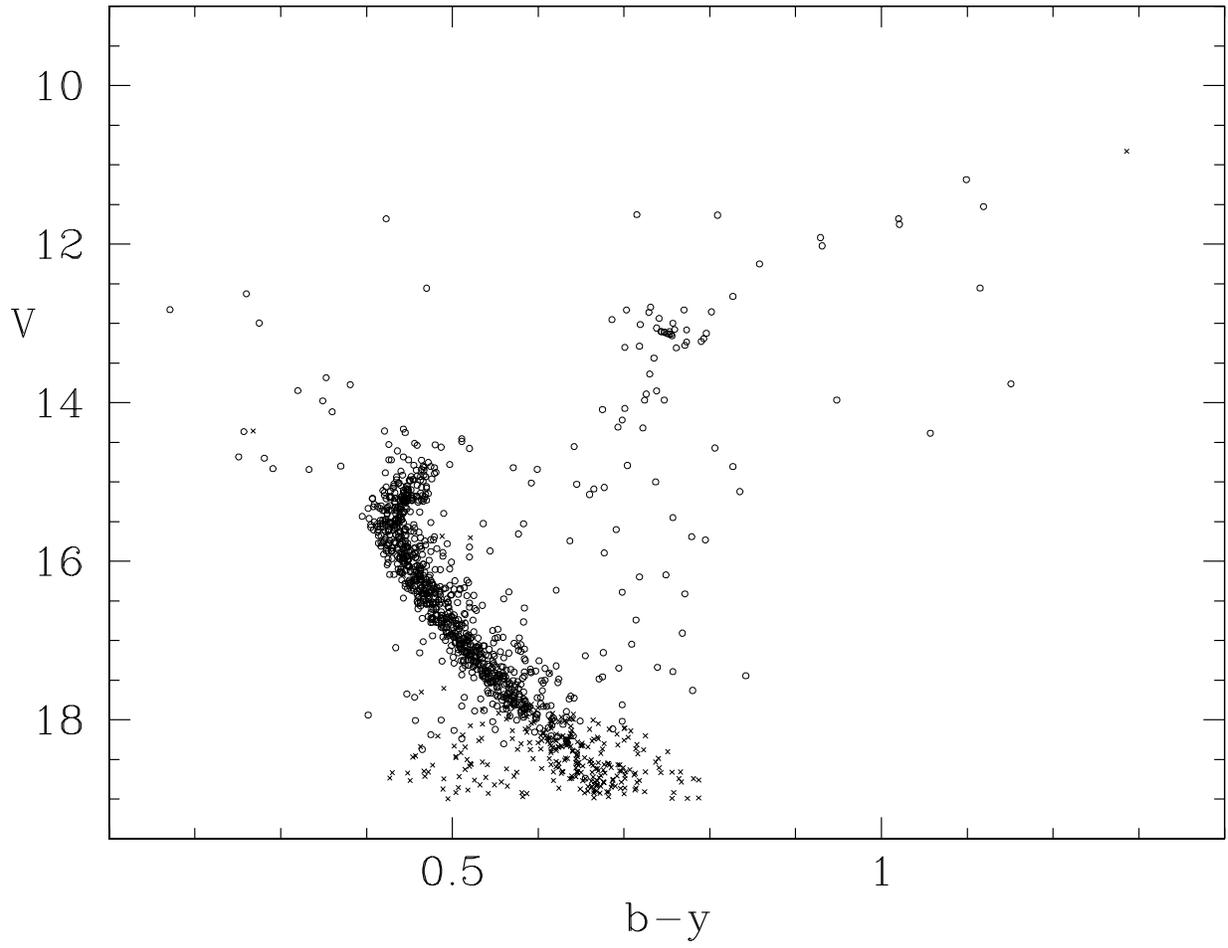}
\caption{Same as Fig. 5 for all stars with proper-motion membership above 50\%.}
\end{figure}

\clearpage
\begin{figure}
\includegraphics[width=\columnwidth,angle=270, scale=0.80]{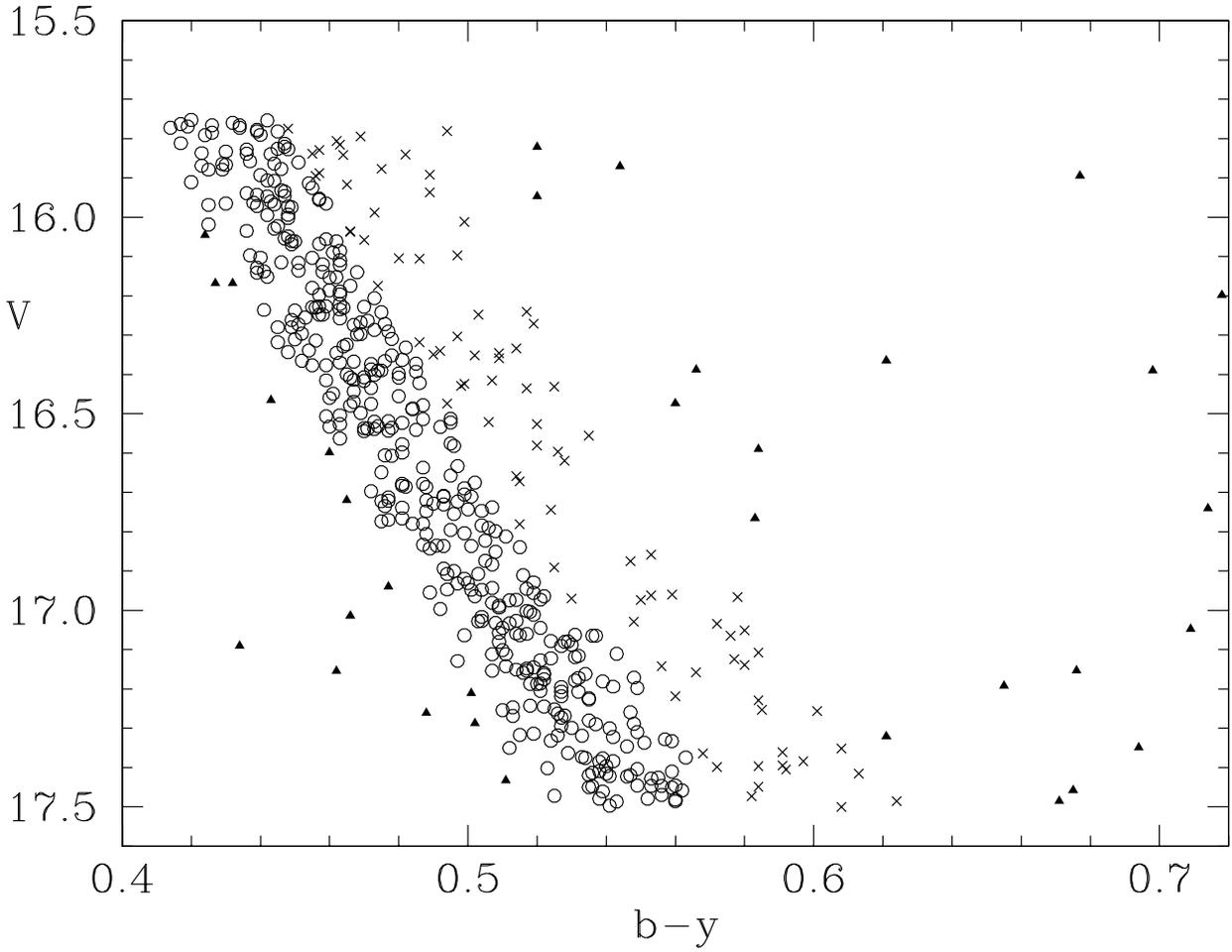}
\caption{The unevolved ($V, b-y$) main sequence of Fig. 7. 
Open circles are probable single stars or binaries with small mass ratios, crosses are potential binaries with q closer to 1.0, and 
filled triangles are likely non-members.}
\end{figure}
\clearpage

\clearpage
\begin{figure}
\includegraphics[width=\columnwidth,angle=270, scale=0.80]{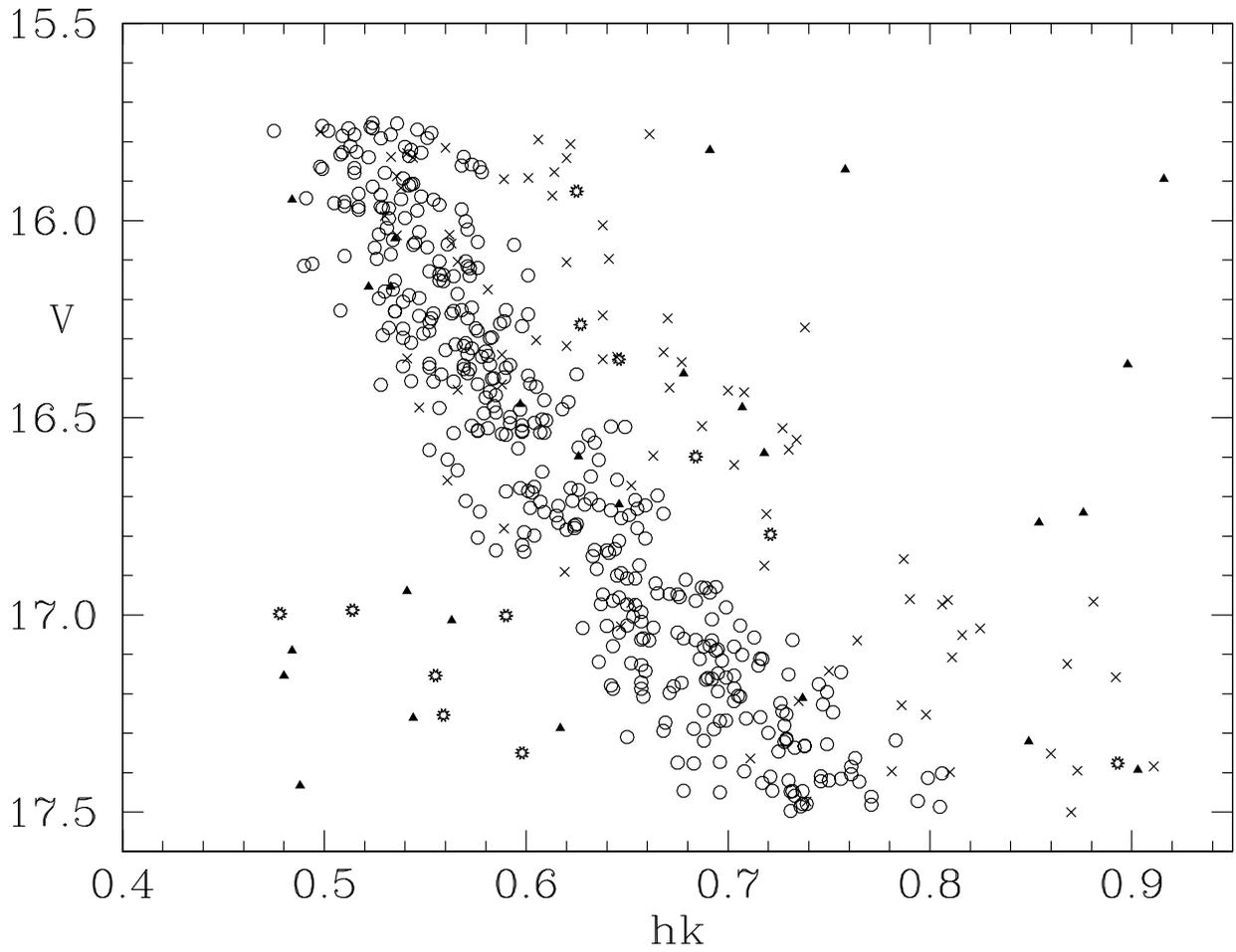}
\caption{The unevolved ($V, hk$) main sequence for stars of Fig. 7. Symbols have the same meaning as for Fig. 8, with the open starburst 
symbols denoting additional deviants for removal.}
\end{figure}
\clearpage

\begin{figure}
\includegraphics[width=\columnwidth,angle=270,scale=0.80]{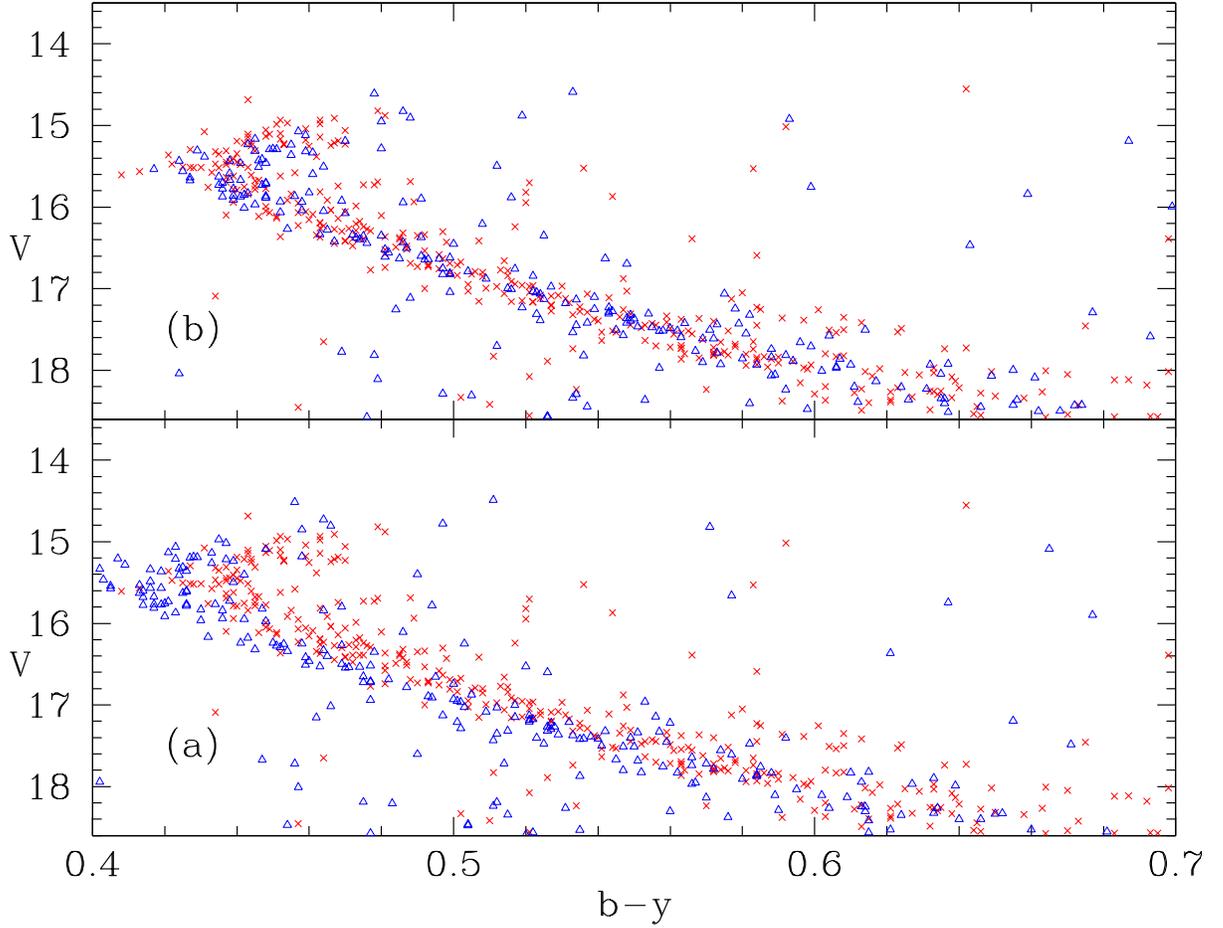}
\caption{(a) The CMD turnoff and main sequence with stars isolated by position on the sky. Red crosses come from regions
of predicted higher reddening by PL while the blue triangles come from the quadrant with the reddening minimum. (b) Same as
(a) with the blue triangles shifted in $b-y$ by +0.022 mag and $V$ by 0.10 mag.}
\end{figure}
\clearpage

\begin{figure}
\includegraphics[width=\columnwidth,angle=270,scale=0.80]{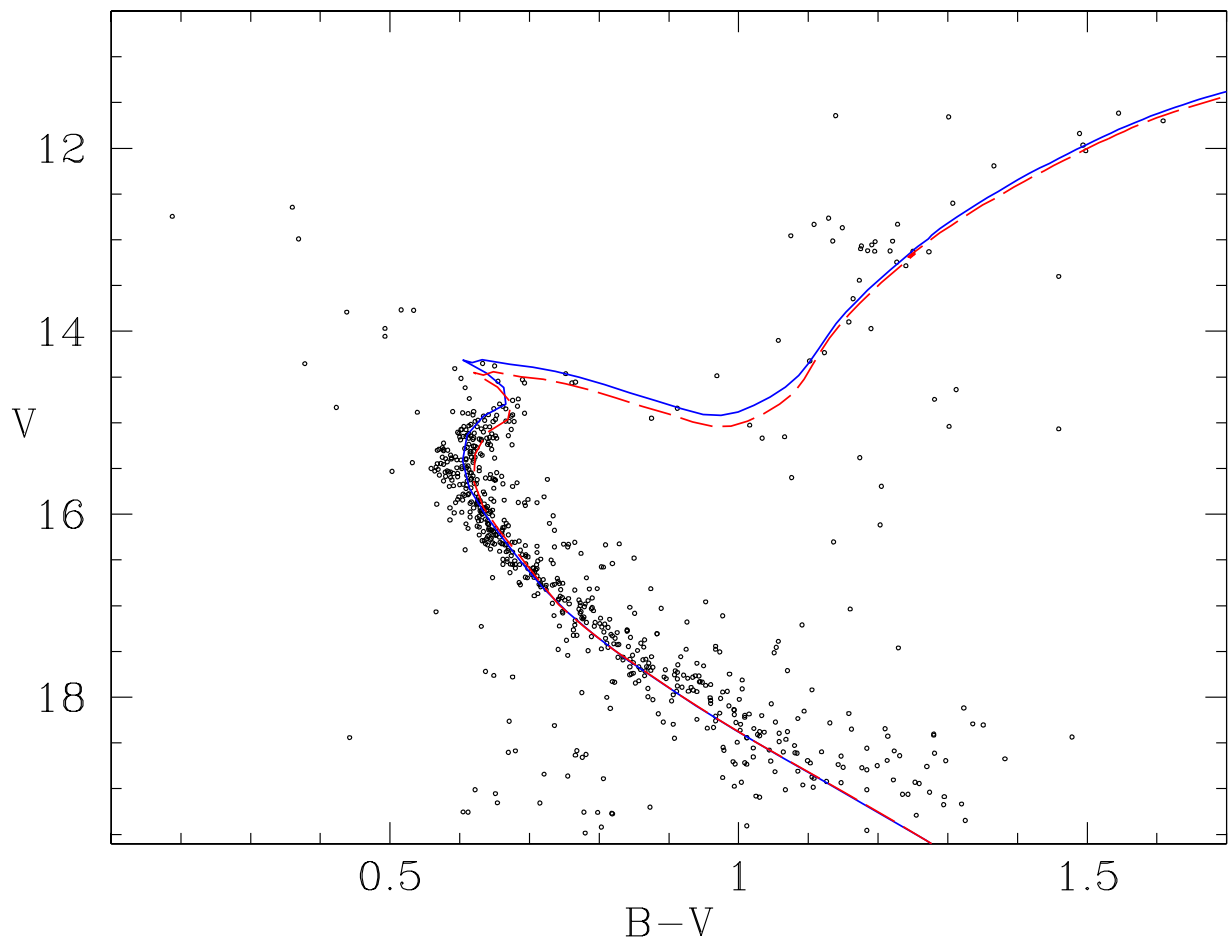}
\caption{The ($V, B-V$) CMD of RV, corrected for the X-dependent trend of Fig. 2, and individual stars corrected for variable reddening 
across the face of the cluster. Superposed $Y^2$ isochrones have [Fe/H] = -0.06 and ages of 2.3 (blue solid curve) and 2.5 Gyr (red dashed curve), adjusted with $E(B-V)$ =
0.16 and $(m-M)$ = 12.40.}
\end{figure}

\clearpage
\input{tab1.tex}
\clearpage
\input{tab2.tex}
\end{document}

%% file: tab1.tex
\begin{deluxetable}{lccccccccc}
\thispagestyle{empty}
\tablecolumns{9}
\tablenum{1}
\tabletypesize\small
\tablewidth{0pc}
\tablecaption{Characterization of Calibration Equations}
\tablehead{
\colhead{Index} & \colhead{Class} & \colhead{$N_{nights}$} & 
\colhead{$N_{pe Stds}$} & \colhead{$\alpha$} & \colhead{$\gamma$} &
\colhead{$\sigma_1$} & \colhead{$\sigma_2$} &  \colhead{sem} }
\startdata
$V$   &\nodata& 4 & 6-15 & 1.0   & 0.045 & 0.027 & 0.024 & 0.005 \cr
$b-y$ & BD/RG & 4 & 4-12 & 0.993 &\nodata& 0.016 & 0.007 & 0.003 \cr
$b-y$ & RD    & 4 & 2-4  & 0.860 &\nodata& 0.048 & 0.006 & 0.004 \cr
$m_1$ & BD    & 3 & 2-4  & 1.0   &\nodata& 0.053 & 0.028 & 0.008 \cr
$m_1$ & RG    & 3 & 2-6  & 1.0   & -0.09 & 0.029 & 0.031 & 0.010 \cr
$m_1$ & RD    & 3 & 2-5  & 1.0   &\nodata& 0.026 & 0.029 & 0.009 \cr
$c_1$ & BD    & 3 & 2-4  & 1.0   &\nodata& 0.031 & 0.017 & 0.035 \cr
$c_1$ & RG    & 3 & 2-6  & 0.95  &  0.25 & 0.144 & 0.042 & 0.050 \cr
$c_1$ & RD    & 3 & 2-4  & 1.0   &\nodata& 0.183 & 0.040 & 0.031 \cr
$hk$  & \nodata   & 2 & 7-13 & 1.057 &\nodata& 0.032 & 0.035 & 0.009 \cr
H$\beta$& \nodata & 1 & 33 & 1.125 &\nodata& 0.017 &\nodata& 0.003 \cr
\enddata
\tablecomments{Calibration equations for index $x$ are of the form $x_{std} = \alpha\  x_{instr} + \gamma (b-y)_{instr} + \beta$. 
Classes of stars include warm dwarfs BD, cool dwarfs RD and cool giants RD. $\sigma_1$ describes greatest dispersion of standard star values
with respect to the calibration equation for any night incorpporated in the calibration; $\sigma_2$ indicates the dispersion among individual nights' calibration equation zeropoints; the sem column denotes the standard error of the mean calibration equation zeropoint value.}
\end{deluxetable}

%% file: tab2.tex
\begin{deluxetable}{rrrrrrrrrrrrrrrrrrrrrr}
\thispagestyle{empty}
\rotate
\tablecolumns{22}
\tablenum{2}
\tabletypesize\tiny
\tablewidth{0pc}
\tablecaption{$uvbyCa$H$\beta$ Photometry of Stars in the Field of NGC 6819}
\tablehead{
\colhead{$\alpha(2000)$} & \colhead{$\delta(2000)$} & \colhead{Prob.} & \colhead{$V$} & 
\colhead{$b-y$} & \colhead{$m_1$} & \colhead{$c_1$} & \colhead{$hk$} & \colhead{H$\beta$} &
\colhead{$\sigma_V$} & \colhead{$\sigma_{by}$} & \colhead{$\sigma_{m1}$} & 
\colhead{$\sigma_{c1}$} & \colhead{$\sigma_{hk}$} & \colhead{$\sigma_{\beta}$} & 
\colhead{$N_y$} & \colhead{$N_b$} & \colhead{$N_v$} & \colhead{$N_u$} & 
\colhead{$N_{Ca}$} & \colhead{$N_n$} & \colhead{$N_w$} }
\startdata 
   295.18539 &  40.30586 &\nodata &   9.532 &  0.088 &  0.156 &  1.077 &  0.309 &  2.843 &  0.004 &  0.010 &  0.014 &  0.012 &  0.014 &  0.014 &   4 &  6 &  8 &  6 &  5 & 10 &  7  \\
   295.11700 &  40.34310 &\nodata &   9.610 &  0.985 &  0.728 & -0.317 &\nodata &  2.518 &  0.001 &  0.002 &  0.006 &  0.009 &  0.018 &  0.015 &   4 &  4 &  2 &  2 &  1 &  9 &  8  \\
   295.48468 &  40.31559 & 3 &   9.660 &  0.371 &  0.177 &  0.397 &  0.646 &  2.580 &  0.002 &  0.004 &  0.008 &  0.010 &  0.006 &  0.009 &   4 &  4 &  7 &  7 & 10 & 10 &  7  \\
   295.17383 &  40.28929 &\nodata &  9.773 &  0.284 &  0.111 &  0.586 &  0.385 &  2.651 &  0.001 &  0.004 &  0.007 &  0.006 &  0.006 &  0.012 &   5 &  7 &  8 &  7 &  8 & 10 &  8  \\
   295.39954 &  40.16243 &\nodata & 10.522 &  1.444 & -1.296 &  1.927 & -0.245 &  2.861 &  0.046 &  0.060 &  0.077 &  0.060 &  0.076 &  0.054 &   4 & 12 & 16 & 12 & 22 & 12 & 14  \\
   295.17725 &  40.05150 & 0 &  10.830 &  0.633 &  0.371 &  0.505 &  1.109 &  2.521 &  0.002 &  0.003 &  0.005 &  0.005 &  0.005 &  0.005 &  10 & 11 & 13 & 14 & 20 & 16 & 16  \\
   295.51831 &  40.24933 & 99 & 10.830 &  1.286 & -0.185 &  0.759 &  0.956 &  2.734 &  0.042 &  0.053 &  0.066 &  0.052 &  0.064 &  0.055 &   9 & 16 & 13 & 14 & 20 & 17 & 16  \\
   295.35254 &  40.05614 & 0 & 10.967 &  0.683 &  0.332 &  0.509 &  1.116 &  2.527 &  0.003 &  0.003 &  0.005 &  0.005 &  0.004 &  0.003 &  13 & 13 & 13 & 14 & 20 & 17 & 15  \\
   295.25128 &  40.23014 & 0 & 10.987 &  0.366 &  0.150 &  0.465 &  0.584 &  2.636 &  0.004 &  0.007 &  0.011 &  0.010 &  0.011 &  0.008 &  18 & 16 & 21 & 14 & 29 & 24 & 23  \\
   295.28415 &  40.32453 &\nodata &  11.181 &  1.063 &  0.690 & -0.293 &  1.790 &  2.611 &  0.006 &  0.008 &  0.010 &  0.010 &  0.009 &  0.009 &  14 & 17 & 13 & 14 & 20 & 17 & 16  \\
  & & & & & & & & & & & & & & & & & & & & \\
   295.47934 &  40.23945 & 99 & 11.187 &  1.099 &  0.624 & -0.271 &  1.718 &  2.625 &  0.005 &  0.007 &  0.011 &  0.015 &  0.011 &  0.009 &  15 & 17 & 13 & 14 & 20 & 17 & 16  \\
   295.32315 &  40.17828 & 0 & 11.361 &  1.202 &  0.106 &\nodata &  1.193 &  2.719 &  0.005 &  0.008 &  0.023 &  0.049 &  0.023 &  0.007 &  23 & 15 &  4 &  1 &  4 & 25 & 24  \\
   295.42532 &  40.30449 & 5 &  11.418 &  0.523 &  0.211 &  0.524 &  0.664 &  2.610 &  0.001 &  0.002 &  0.004 &  0.005 &  0.003 &  0.004 &  15 & 16 & 13 & 14 & 20 & 17 & 16  \\
   295.39221 &  40.05499 & 0 & 11.473 &  0.321 &  0.124 &  0.485 &  0.439 &  2.651 &  0.002 &  0.003 &  0.005 &  0.005 &  0.004 &  0.005 &  17 & 15 & 13 & 14 & 20 & 17 & 16  \\
   295.29922 &  40.22464 & 99 & 11.528 &  1.119 &  0.524 &  0.257 &  1.801 &  2.605 &  0.002 &  0.003 &  0.006 &  0.010 &  0.005 &  0.005 &  25 & 25 & 22 & 14 & 29 & 25 & 24  \\
   295.29578 &  40.18643 & 99 & 11.629 &  0.715 &  0.361 &  0.526 &  1.126 &  2.571 &  0.003 &  0.003 &  0.006 &  0.007 &  0.005 &  0.005 &  24 & 25 & 22 & 14 & 30 & 25 & 24  \\
   295.29630 &  40.19488 & 99 & 11.636 &  0.809 &  0.495 &  0.417 &  1.327 &  2.559 &  0.002 &  0.003 &  0.005 &  0.006 &  0.004 &  0.004 &  24 & 22 & 22 & 14 & 30 & 25 & 20  \\
   295.39288 &  40.29617 & 99 & 11.647 &  0.978 &  0.698 &  0.172 &  1.790 &  2.536 &  0.002 &  0.003 &  0.005 &  0.006 &  0.005 &  0.005 &  17 & 17 & 13 & 14 & 21 & 19 & 17  \\
   295.32086 &  40.18101 & 99 & 11.680 &  1.020 &  0.652 &  0.140 &  1.764 &  2.572 &  0.002 &  0.003 &  0.005 &  0.008 &  0.006 &  0.004 &  25 & 25 & 22 & 13 & 26 & 25 & 24  \\
   295.21530 &  40.11550 & 99 & 11.681 &  0.423 &  0.182 &  0.398 &  0.697 &  2.602 &  0.004 &  0.006 &  0.010 &  0.009 &  0.008 &  0.005 &  24 & 20 & 22 & 14 & 30 & 25 & 24  \\
\enddata
\end{deluxetable}